\documentclass[conference, 10 pt]{IEEEtran} 		% For final IEEE paper version
\IEEEoverridecommandlockouts
\usepackage{ushort}
\usepackage[mathcal]{euscript}
\usepackage{amsmath}
\usepackage{amsthm}
\usepackage{amssymb}
\usepackage{graphicx}
\usepackage{epstopdf}
\usepackage{booktabs}% http://ctan.org/pkg/booktabs
\usepackage{colortbl}% http://ctan.org/pkg/colortbl
\usepackage{amsmath}% http://ctan.org/pkg/amsmath
\usepackage{xcolor}% http://ctan.org/pkg/xcolor
\usepackage{graphicx}% http://ctan.org/pkg/graphicx
\usepackage{algpseudocode}
\usepackage{algorithmicx}
\usepackage{algorithm}
\algnewcommand{\algorithmicgoto}{\textbf{go to}}%
\algnewcommand{\Goto}[1]{\algorithmicgoto~\ref{#1}}%
\usepackage{cite}
\usepackage{rotating}
\usepackage{array}
\def \aed {\lambda_{\text{ed}}}

\def \bedi {\beta_{\text{ed}}^i}
\def \bpui {\beta_{\text{pu}}^i}

\def \med {\mu_{\text{ed}}}
\def \mpu {\mu_{\text{pu}}}

\def \ued {U_{\text{ed}}}
\def \upu {U_{\text{pu}}}

\begin{document}

\title{A Delay Efficient Multiclass Packet Scheduler for Heterogeneous M2M Uplink}

\author{\IEEEauthorblockN{Akshay Kumar, Ahmed Abdelhadi and Charles Clancy}
\IEEEauthorblockA{Hume Center, Virginia Tech\\
Email:\{akshay2, aabdelhadi, tcc\}@vt.edu}
\thanks{This research is based upon work supported by the National Science Foundation under Grant No. 1134843. 
}
}

\maketitle 
\thispagestyle{plain}
\pagestyle{plain}

\begin{abstract}
The sensory traffic in Machine-to-Machine (M2M) communications has fairly heterogeneous service delay requirements. Therefore, we study the delay-performance of a heterogeneous M2M uplink from the sensors to a M2M application server (AS) via M2M aggregators (MA). We classify the heterogeneous M2M traffic aggregated at AS into multiple Periodic Update (PU) and Event Driven (ED) classes. The PU arrivals are periodic and need to be processed by a prespecified \emph{firm} service deadline whereas the ED arrivals are random with \emph{firm} or \emph{soft} real-time or non real-time service requirements. We use step and sigmoidal functions to represent the service utility for PU and ED packets respectively. We propose a delay efficient multiclass packet scheduling heuristic that aims to maximize a proportionally fair system utility metric. Specifically, the proposed scheduler prioritizes service to ED data while ensuring that the PU packets meet their service deadline. It also minimizes successive PU failures for critical applications by penalizing their occurrences. Furthermore, the failed PU packets are immediately cleared from the system so as to reduce network congestion. Using extensive simulations, we show that the proposed scheduler outperforms popular packet schedulers and the performance gap increases with heterogeneity in latency requirements and with greater penalty for PU failures in critical applications.
\end{abstract}

\begin{IEEEkeywords}
M2M, Latency, Quality-of-Service, Multiclass Scheduler
\end{IEEEkeywords}

\section{Introduction}
Machine-to-Machine (M2M) communications is becoming increasingly ubiquitous due to its vast number of residential and industrial applications such as in smart homes, vehicle tracking, industrial automation etc. In most of the M2M applications, the uplink traffic generated from the sensors is fairly heterogeneous and can be classified as either non real-time (no deadline for task completion) or \emph{soft} real-time (decreased utility if deadline not met) or \emph{firm} (zero utility if deadline not met). For instance, the messages from a instrumented protective system have \emph{firm} real-time service requirements whereas preventive maintenance applications are typically served in non real-time. With increase in network size, the available computational and communication resources gets shared among a large number of sensors which makes it very hard to provide real-time service \cite{3GPPreport}. This necessitates the need of designing delay-efficient packet scheduler for heterogeneous M2M uplink and that is applicable for any M2M application, communication standard and hardware-software architecture. 

Most of the existing M2M packet schedulers are designed for specific wireless technology such as LTE (see \cite{Gotsis12} and references therein) and fail to fully characterize the heterogeneity in M2M uplink traffic as done in our work. Another line of work considers the design of packet schedulers for real-time control applications (see \cite{Buttazzo11} and references therein) that guarantee the schedulability of critical tasks while providing best-effort service for others. These works assume that the relative deadline for periodic tasks is equal to their periods, all aperiodic tasks are aggregated into a single class and the minimum inter-arrival time of an aperiodic task equals its deadline. In our work, we do not make any of these restrictive assumptions on the nature of M2M traffic. Also, unlike these works, the objective of our scheduler is to ensure utility-proportional fairness among all traffic classes.  

The work that is most closely related to our work is by Kumar et. al. in \cite{sysCon16} wherein a delay-efficient heuristic scheduler is proposed for M2M uplink. However, this work does not account for the diverse delay requirements of traffic from different sensors and it classifies the aggregated sensor traffic into two broad PU and ED classes. Also, it does not attempts to minimize successive PU failures, which may result in undesired consequences for the M2M system.

In this paper, we develop a delay-efficient mutliclass packet scheduler for heterogeneous M2M uplink traffic. To begin with, we use source M2M traffic model in \cite{Navid13} to classify the data from each sensor as Periodic Update (PU) or Event Driven (ED). The PU arrivals are periodic with \emph{firm} service deadline while the ED arrivals are random with \emph{firm} or \emph{soft} real-time or non real-time service requirements. We map the contrasting delay requirements of PU and ED packets onto step and sigmoidal utility functions respectively. The goal of proposed scheduler is to maximally satisfy the delay requirement of all classes which translates to maximizing a proportionally-fair system utility metric. It does so by prioritizing service to ED packets while ensuring that the PU packets meet their deadline. This results in a higher utility for the ED packets without compromising the utility for PU packets. 

However, we note that, due to random arrival and service times, some PU packets fail to meet their deadline. It becomes particularly difficult to do so as congestion increases with increase in network size. When this occurs, we remove failed PU packets to reduce network congestion and improve overall system utility. Furthermore, for critical M2M applications, successive PU failures are highly detrimental to system performance. Hence, we aim to minimize successive PU failures by imposing a penalty to the PU utility. Using extensive simulations, we show that the proposed scheduler has a superior delay-performance compared to popular scheduling policies such as First-Come-First-Serve (FCFS), Earliest Due Date (EDD), fixed priority scheduler etc. The performance gap increases with heterogeneity in latency requirements and with higher penalty for successive PU failures for critical applications. Another salient feature of the proposed scheduler is that it is general enough to be used across different M2M applications, M2M communication standards and hardware-software M2M architectures. 

 The rest of the paper is organized as follows. Section~\ref{sysModel} introduces the system model and defines the utility functions for PU and ED data. Then in Section~\ref{probForm}, we formulate the utility maximization problem and in Section \ref{propSch} describe the proposed packet scheduling heuristic. Section~\ref{Results} presents simulation results. Finally Section~\ref{concl} draws some conclusions.

\section{System Model}
\label{sysModel}
\begin{figure}%
\centering
\includegraphics[width=\columnwidth]{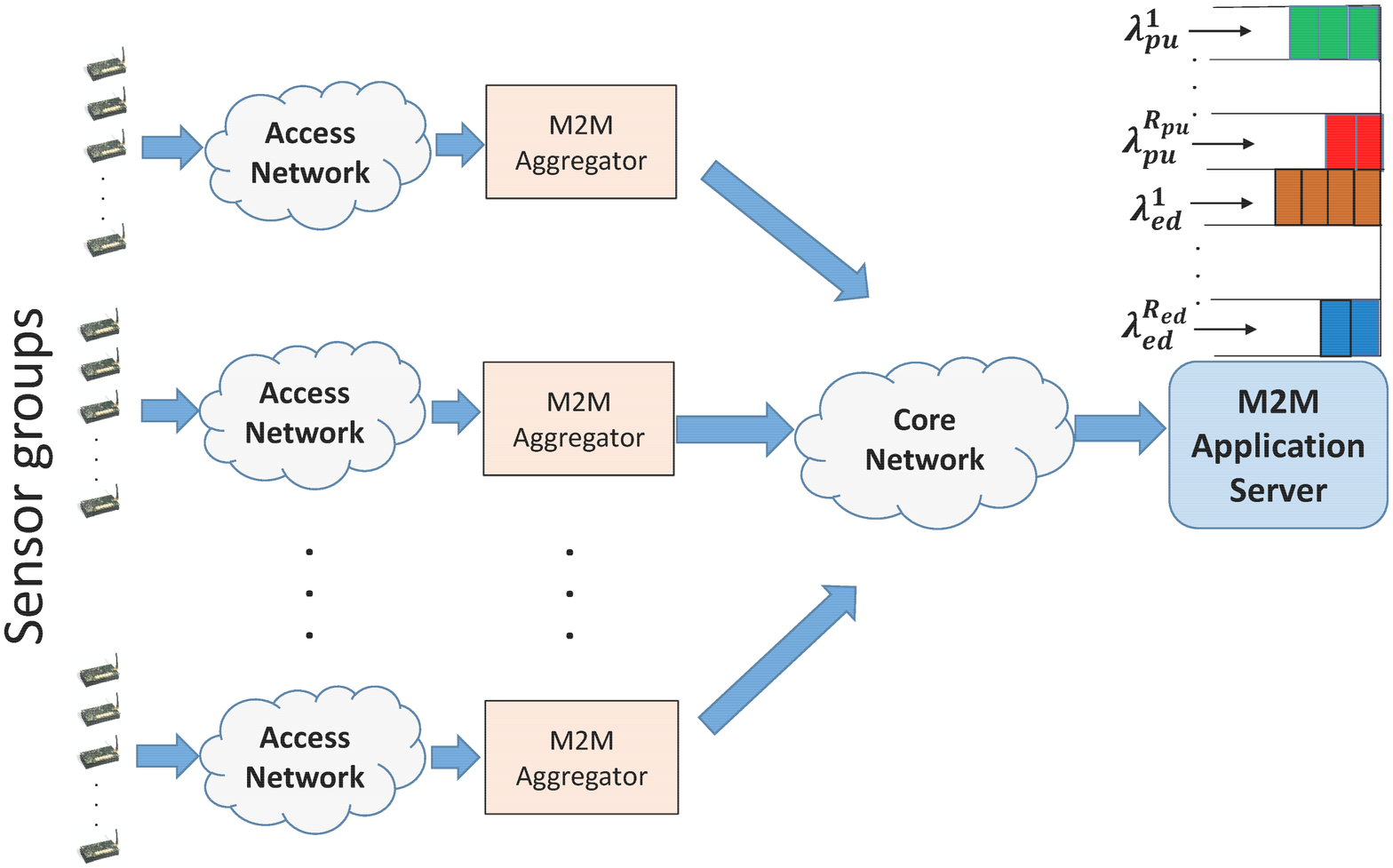}%
\vspace{0pt}
\caption{System model illustrating queuing process at M2M AS.}%
\label{systemModel}%
\end{figure}

Fig~\ref{systemModel} shows the system model for a heterogeneous M2M uplink from $N$ sensors to a M2M application server (AS) via multiple M2M aggregators. The aggregated M2M traffic at AS is classified into $R_{\text{pu}}$ PU and $R_{\text{ed}}$ ED classes. Here, we assume that there exists multiple sensors with similar arrival period and service deadlines for PU packets and therefore the PU traffic at AS can be clustered into $R_{\text{pu}}$ classes. Similarly we cluster the aggregated ED traffic into $R_{\text{ed}}$ classes based on the closeness in the characteristics of ED traffic. The period and relative deadline for $i^{\text{th}}$ PU class are $T_{\text{pu}}^i$ and $l_d^i$ respectively. The ED arrivals is modeled as a Poisson process with rate $\aed^i$ for $i^{\text{th}}$ class. 

%\vspace{-10pt}
The service times for $i^{\text{th}}$ PU and ED class are assumed to be exponential with rates $\mpu^i$ and $\med^i$ respectively. In this work, we consider only queuing and service delay at AS and ignore other components. Due to relatively small payload for M2M traffic, the packet transmission time are small relative to queuing delay and thus can be ignored. We assume sufficient number of wireless resources for both access and core network to ignore the congestion delay. Now the scheduling policy at AS should be chosen as to maximally satisfy the delay requirements of each PU and ED class. 

\section{Problem Formulation} 
\label{probForm}

\subsection{Utility Functions}
\subsubsection{PU utility} 
\begin{figure} 
\centering
   \includegraphics[width=\columnwidth]{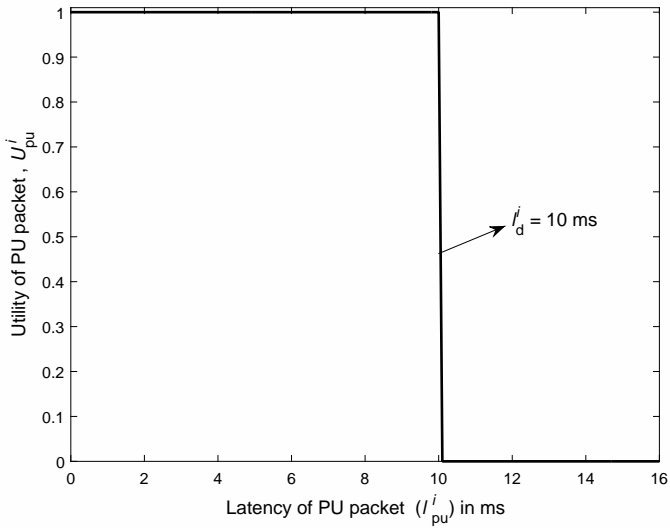}
\caption{\footnotesize{Utility function for PU packet of class $i$.}}
\label{utilFuncPU}
\end{figure}

As stated previously, the PU packets need to be processed within a prespecified time interval at the end of which there is no utility in serving the packet. For a PU packet of $i^{\text{th}}$ class with latency $l_{\text{pu}}^i$, we define the utility function (see Fig.~\ref{utilFuncPU}) as,
\begin{equation}
\upu^i(l_{\text{pu}}^i) = \begin{cases} 1 & \text{if } l_{pu}^i < l_d^i \\
 0 & \text{if } l_{pu}^i \geq l_d^i, \end{cases}
\label{puUtila}
\end{equation}
where $l_d^i$ is the latency deadline at which utility drops to 0. 

We next define a penalty function to limit the number of successive PU failures. Let $r_i$ denote the run-length\footnote{A run is a maximal length sequence of identical outcomes within a larger sequence. For example, let $[S S {\color{red}F F} S {\color{red} F} S {\color{red}F F F} S]$ denote a sequence of PU success ($S$) and failure (${\color{red}F}$). Here, we have three runs of PU failures with length $r_i$ equal to $1$, $2$ and $3$ respectively.} of a sequence of PU failures for $i^{\text{th}}$ class, then the penalty for the run (besides the utility loss for each PU failure) is,
\begin{equation}
P_{\text{pu}}^i(r_i) = r_i - {r_i}^{\gamma_i},~r_i \geq 1
\label{puUtilb}
\end{equation}
where $\gamma_i \geq 1$ indicates the severity of the penalty for $i^{\text{th}}$ class and is chosen based on the criticality of the application. For instance, consider the smart-grid uplink traffic which besides other messages, consists of periodic real-time pricing (RTP) updates and periodic meter readings \cite{openSG}. The RTP updates are highly critical messages and failure to process them within a particular time can potentially result in significant economic loss \cite{rtpUpdate}. On the contrary, failure to process periodic meter readings in time is not a big concern. It can easily be estimated using the meter reading of last successfully received (at the M2M application server) meter reading and any differences in the estimate and actual reading can be adjusted in future billing cycle. Therefore, the penalty ($\gamma_i$) for RTP failures should be higher than that for meter readings. The special case of $\gamma_i = 1$ corresponds to when there is no penalty for successive PU failures.

\subsubsection{ED utility} 
\begin{figure}
\centering
\includegraphics[width=\columnwidth]{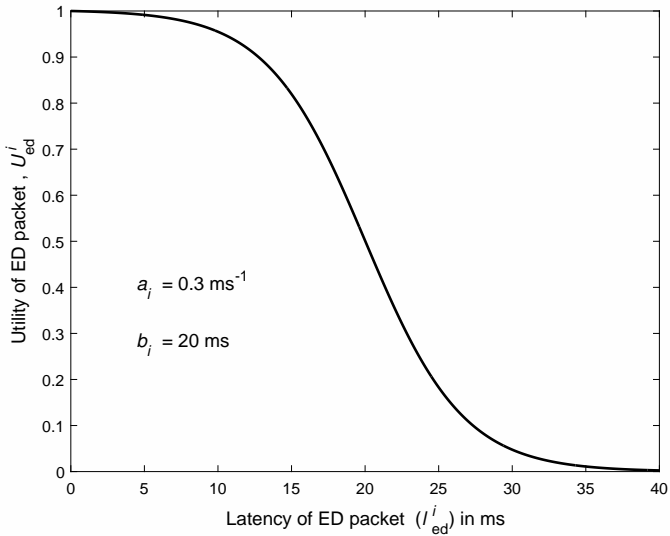}%
\caption{\footnotesize{Utility function for ED packet of class $i$.}}
\label{utilFuncED} 
\end{figure}

Based on the M2M application, the delay requirements of ED traffic may vary from \emph{firm} to non real-time service requirements. Sigmodial functions are fairly versatile to model diverse Quality-of-Service requirements by appropriately varying their parameters \cite{AbdelhadiCNC2014, AbdelhadiPIMRC2013}. Therefore, we use sigmoidal function to model the utility of $i^{\text{th}}$ ED class (see Fig.~\ref{utilFuncED}) as,
\begin{equation}
\ued^i(l_{\text{ed}}^i) = 1-c_i\left(\frac{1}{1+{\text{e}}^{-a_i(l_{\text{ed}}^i-b_i)}}-d_i\right),
\label{edUtil}
\end{equation}
where, $c_i=\frac{1+{\text{e}}^{a_ib_i}}{{\text{e}^{a_ib_i}}}$ and $d_i = \frac{1}{1+{\text{e}^{a_ib_i}}}$. The parameter $a_i$ is the utility roll-off factor that signifies the criticality of the application while the parameter $b_i$ roughly indicates a soft latency deadline.

\subsection{System utility function} 
%For a given scheduling policy $\mathcal{P}$, we first define a proportionally fair system utility function as,
%\begin{equation}
%V(\mathcal{P}) = {\bold{\upu}}^{\bpu}(\mathcal{P})*{\bold{\ued}}^{\bed}(\mathcal{P}),
%\label{sysUtil}
%\end{equation}
%where $\bold{U_i}(\mathcal{P}) = U_i(\tilde{l}_i)$ is the utility for class $i$ in steady state.
For a given scheduling policy $\mathcal{P}$, let $\hat{U}_{\text{pu}}^i(\mathcal{P})$ and $\hat{U}_{\text{ed}}^i(\mathcal{P})$ denote the average utility of $i^{\text{th}}$ PU and ED class respectively, as given by,
\begin{align}
\hat{U}_{\text{pu}}^i(\mathcal{P}) &= \lim_{T_s \rightarrow \infty} \frac{\sum_{j=1}^{M_{\text{pu}}^i(T_s)} U_{\text{pu}}^i(l_{\text{pu}}^{ij}) + \sum_{j=1}^{R_i(T_s)} P_{\text{pu}}^i(r_{ij})}{M_{\text{pu}}^i(T_s)}, \\
\hat{U}_{\text{ed}}^i(\mathcal{P}) &= \lim_{T_s \rightarrow \infty} \frac{\sum_{j=1}^{M_{\text{ed}}^i(T_s)} U_{\text{ed}}^i(l_{\text{ed}}^{ij})}{M_{\text{ed}}^i(T_s)},
\end{align}
where $M_{\text{pu}}^i(T_s)$ and $M_{\text{ed}}^i(T_s)$ are the number of packets from $i^{\text{th}}$ PU and ED class served in time $T_s$. The variable $r_{ij}$ denotes the length of $j^{\text{th}}$ sequence or run of PU failures for class $i$ and $R_{i}(T_s)$ denotes the number of PU failure sequences in time $T_s$. Lastly, $l_{\text{pu}}^{ij}$ and $l_{\text{ed}}^{ij}$ denote the latency of the $j^{\text{th}}$ packet $i^{\text{th}}$ PU and ED class respectively. 

We next aggregate the average PU and ED utilities into the following proportionally-fair system utility metric.
\begin{equation}
V(\mathcal{P}) = \prod_{i=1}^{R_{\text{pu}}} {(\hat{U}_{\text{pu}}^i)}^{\bpui}(\mathcal{P}) * \prod_{i=1}^{R_{\text{ed}}}{(\hat{U}_{\text{ed}}^i)}^{\bedi}(\mathcal{P}),
\label{sysUtil}
\end{equation}
where the parameters $\bpui$ and $\bedi$ denote the relative priority assigned to $i^{\text{th}}$ PU and ED class respectively.

\subsection{Delay-optimal scheduling policy} Our objective is to determine the optimal scheduling policy $\mathcal{P}$ at M2M AS that maximizes the proposed system utility metric, $V(\mathcal{P})$. If the arrival time and service times of packets of each PU and ED class are known \emph{apriori}, then the optimal scheme is to process packets of $i^{\text{th}}$ PU class such that their service is completed exactly at their deadline. However, due to random ED arrivals and random service times, it is not possible to determine an online optimal scheduler \cite{TiaLiuShankar}. Therefore, we propose an online heuristic scheduler with a goal to maximize system utility, as described in the next section.

\section{Proposed Scheduler}
The proposed scheduler dynamically schedules packets of different classes using delay-efficient heuristics that aim to maximize the proposed system utility metric. Therefore, the backlogged packets at the AS may preempt the packet in service depending on the following scheduling heuristics.
\label{propSch}
\subsection{Service order between PU and ED classes}
From the PU utility function in Fig.~\ref{utilFuncPU}, it is clear that there is no added utility of processing it before the deadline $l_d^i$. On the contrary, the ED utility decreases monotonically with increase in latency. Therefore, the proposed scheduler should prioritize service to backlogged ED packets until the delay for PU packet of class $i$ exceeds certain delay threshold $l_t^i (l_t^i<l_d^i)$. Note that, due to randomness in arrival and service times of packets, it is not possible to guarantee that all PU packets will meet their deadline. However, the value of $l_t^i$ is chosen such that the .

\subsection{Service order among PU classes}
For resolving service contention between packets of different PU classes, we give higher (preemptive) priority to the PU class with higher penalty factor $\gamma_i$ and lower service rate $\mu_{\text{pu}}^i$. This is because higher $\gamma_i$ for PU class indicates that it is less robust to PU failures and therefore, should be served first. We prioritize service to PU class with lower service rate (i.e., higher average service time) because it is more likely to be unsuccessful if its service is delayed. Again due to the \emph{firm} utility function for PU classes, if there are backlogged ED packets, then the priority order between two PU classes comes into effect only when packets of each PU class exceeds its corresponding latency threshold.%When a PU class  has higher $\gamma_i$ and higher $\mu_{\text{pu}}^i$ compared to other, we assign higher priority to class with higher $\gamma_i$ as it is the dominant factor that affects PU utility.

\subsection{Service order among ED classes}
We now discuss how the proposed scheduler determines the order of service between packets of different ED classes. We assign higher priority to ED classes that are more delay sensitive (i.e., have higher $a_i$ and/or lower $b_i$). Unlike the earlier heuristics, the (preemptive) priority order among ED classes comes into effect immediately at the arrival of ED packet. However, if the latency of $i^{\text{th}}$ ED class exceeds a certain threshold $\delta_i$ while there are backlogged packet of lower priority ED classes, then its gets preempted by the ED class with the highest priority among lower priority (backlogged) ED classes. By definition, the threshold $\delta_i = \infty$ for ED class with least priority. 

The preemption policy for ED classes exceeding their threshold $\delta_i$ ensures that ED packets with unusually large service time do not hog the server which will greatly increase the latency for other ED classes. Thus, the proposed heuristic aims to reduce the latency of delay-sensitive ED class without incurring very large latency for delay-tolerant classes. 

\subsection{Service order among packets of same class}
We serve the packets of a given PU or ED class using FCFS policy. Although the Shortest Processing Time (SPT) algorithm is optimal for minimizing the average latency \cite{sptOptimal}, we cannot implement it here because we assume random service time for packets and hence this information is not known \emph{apriori} to the scheduler. Since the average service time is same for all packets of a particular class, FCFS is the best policy in this case.

\subsection{Complexity Analysis}
For the proposed scheduler to maximize the system utility, we need to determine the jointly optimal value of PU and ED thresholds i.e., $l_t^i, \delta_i$. However, the computational complexity for this optimization may be huge if we perform a brute force search over all permissible values of thresholds $l_t^i$ and $\delta_i$, particularly for large number of PU and ED classes. Therefore, we next present the heuristics to reduce the search space for optimization, without sacrificing much on the delay-performance. 

\emph{Search space for the PU threshold:} The unconstrained search space for the PU threshold for class $i$ is $l_t^i \in [0, l_d^i]$. We have assumed the service time for PU packets of class $i$ to be exponentially distributed with rate $\mu_{\text{pu}}^i$. So, the $99$ percentile for service time is roughly $t_{99}^i = 4.6/\mu_{\text{pu}}^i$. Therefore, we reduce the search space for class $i$ to $l_t^i \in [l_d^i-t_{99}^i, l_d^i]$ without significantly affecting its delay-performance. The computational savings are significant for PU class with large deadline $(l_d^i)$ and high service rate $(\mu_{\text{pu}}^i)$. 

\emph{Search space for the ED threshold:} The unconstrained search space for the $i^{\text{th}}$ ED class threshold is $\delta_t^i \in [0, \infty]$. Let $l_{\text{ed},x}^i$ denote the latency at which the utility of packet of $i^{\text{th}}$ ED class equals $x$ and can be easily determined using \eqref{edUtil}. We set the upper bound on $\delta_t^i$ to $l_{\text{ed},0.01}^i$. Since a class $i$ packet on average takes $t_{\text{av}}^i = 1/\mu_{\text{pu}}^i$ time at server, a reasonable lower bound for $\delta_t^i$ is $\text{min}(0, l_{\text{ed},0.99}^i-t_{\text{av}}^i)$. Hence, the reduced search space for class $i$ is $\delta_t^i \in [\text{min}(0, l_{\text{ed},0.99}^i-t_{\text{av}}^i), l_{\text{ed},0.01}^i]$.
 
Lastly, the proposed scheduler drops the backlogged PU packets that have exceeded their latency deadline as there utility drops to $0$. However, this helps clear congestion and thus reduce latency for other backlogged packets. The proposed scheduler is described in detail in Algorithm~\ref{propScheduler}.
 
 \begin{algorithm}
\caption{Proposed Mutliclass Packet Scheduler}
\label{propScheduler}
	\textbf{Begin Algorithm:}
		\begin{algorithmic} 
		\label{algoProp}
		\While{there are backlogged packets} 
		%%%%%%%%%%%%%%%%%%%%%%%%%	
		   \State Remove the failed PU packets. 
        	\State Find classes with packet delay exceeding threshold.
			\If{some PU classes exceed their threshold}
				\State Serve the PU class with highest priority among PU class that exceed their threshold.		
        	\ElsIf{no PU or ED class exceed its threshold}
        		\State Serve the ED class with overall highest priority. 
        	\ElsIf{no PU but some ED classes exceed their threshold}
				\State Serve the ED class with highest priority among ED classes that do not exceed their threshold.         	
            \Else 
             \State Continue processing the current packet in server. 
             \EndIf
		\EndWhile
	\end{algorithmic}	
\end{algorithm}

\section{Simulation Results}
\label{Results}
In this section, we use Monte-Carlo simulations to evaluate the system utility performance of our scheduler against various standard scheduling policies such as FCFS, EDD, Preemptive priority scheduling. The simulation time $T_s$ is set to a large value ($40$~s) to ensure a steady-state queuing behavior. Although, the proposed scheduler is designed for any value of $R_{\text{pu}}$ and $R_{\text{ed}}$, we set $R_{\text{pu}} = R_{\text{ed}} = 2$ to gain practical insights about its performance. The total number of sensors is $N=500$. Of these, $300$ sensors transmit packets of PU class $1$ and the remaining $200$ transmit packets of PU class $2$. For ED traffic, we assume $150$ sensors for class $1$ and the remaining $350$ for class $2$. The PU period for each sensor of type class $1$ and $2$ are $T_{\text{pu}}^1 = 1.8$~s and $T_{\text{pu}}^2 = 0.5$~s respectively. The overall arrival rate for ED class $2$ is $\lambda_{\text{ed}}^2=0.1$~packet/ms whereas $\lambda_{\text{ed}}^1$ is varied from $0.1$--$0.25$~packet/ms. The service rates for each class is set to $1$~packet/ms. 

All classes are assigned equal priority, unless mentioned otherwise. The latency deadline for PU class $1$ and $2$ are $l_d^1 = 4$~ms and $l_d^2=8$~ms. The utility function parameters for ED class $1$ and $2$ are set to $[a_1, b_1] = [1, 10]$ and $[a_2, b_2] = [0.7, 20]$ respectively, unless mentioned otherwise. The penalty parameters $\gamma_1=\gamma_2=1$ unless mentioned otherwise. We initially fix the latency threshold for ED class $1$ and $2$ to $\delta_1 = b_1+(4/a_1)$ and $\delta_2 = \infty$ respectively and later study the performance improvement by optimizing $\delta_1$ over $[\text{min}(0, l_{\text{ed},0.99}^1-t_{\text{av}}^1), l_{\text{ed},0.01}^1]$, where $t_{\text{av}}^1 = 1$~ms in our simulations. For each simulation scenario, we determine the optimal latency threshold for $i^{\text{th}}$ PU class $l_t^i$ by searching over $[l_d^i-t_{99}^i, l_d^i]$. 

\subsection{Impact of delay heterogeneity}
\begin{figure}[!h]%
\centering
\includegraphics[scale = 0.5]{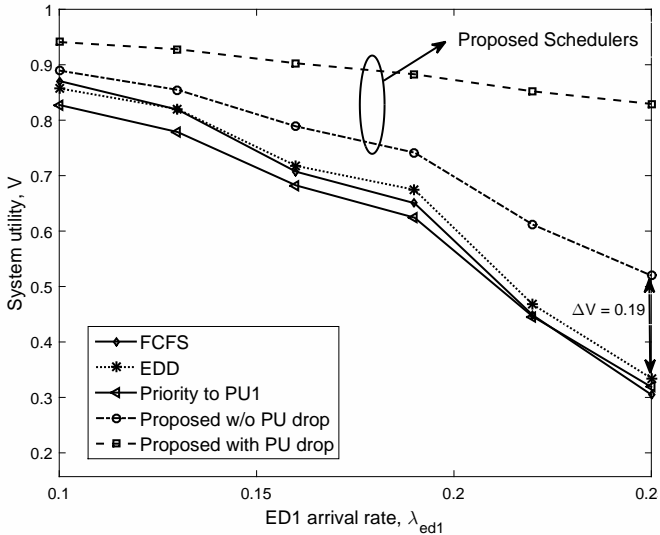}%
\caption{System utility with delay-homogeneous M2M traffic.}%
\label{vary_ed1_homo}%
\end{figure}

\begin{figure}[!h]%
\centering
\includegraphics[scale = 0.5]{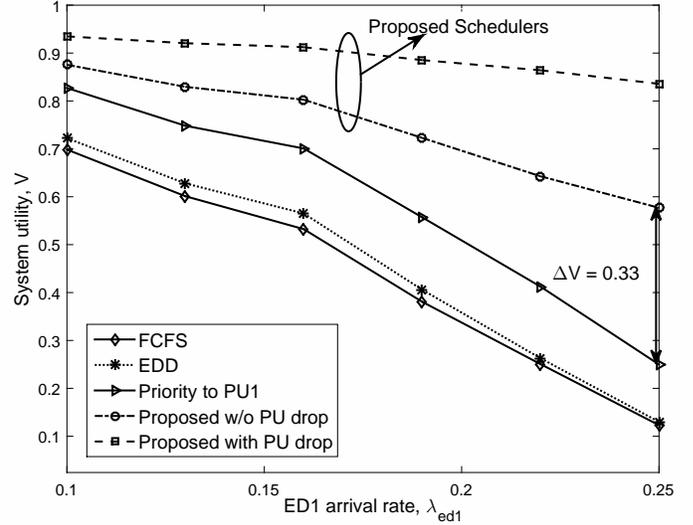}%
\caption{System utility with delay-heterogeneous M2M traffic.}%
\label{vary_ed1_noPenalty}%
\end{figure}

 Fig.~\ref{vary_ed1_homo} shows plot of system utility for a nearly homogeneous system with $T_{\text{pu}}^1 = T_{\text{pu}}^2 = 0.8$~s, $[l_d^1, l_d^2] = [7.8, 8]$~ms, $[a_1, a_2] = [0.65, 7]~{\text{ms}}^{-1}$ and $[b_1, b_2] = [19, 20]$~ms. We then increase the heterogeneity in system by using the default parameters which results in system utility as shown in Fig.~\ref{vary_ed1_noPenalty}. We note that the performance gap between the proposed scheduler and the other schemes, $\Delta V$, is much larger for heterogeneous system ($\Delta V=0.33$) compared to homogeneous system ($\Delta V=0.19$), especially at higher $\lambda_{\text{ed}}^1$. This is because in the proposed scheduler, the effect of priority order among classes on the system utility becomes more pronounced with increasing heterogeneity in classes. Lastly, the performance of proposed scheduler further improves when we drop the failed PU packets due to reduced congestion.

\subsection{Impact of choice of ED threshold}
\begin{figure}[!h]%
\centering
\includegraphics[scale = 0.5]{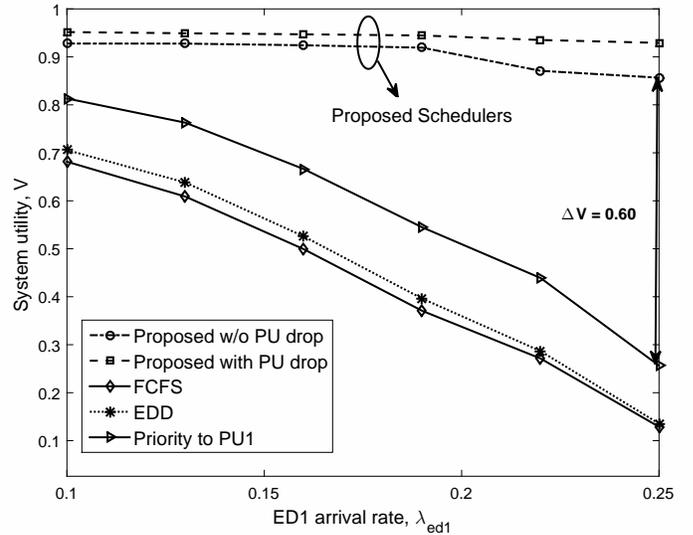}%
\caption{Impact of optimal latency threshold for ED class $1$.}%
\label{vary_ed1_optDelta}%
\end{figure}

Fig.~\ref{vary_ed1_optDelta} shows the impact of selecting optimal threshold $\delta_1$ for ED class $1$ on the system utility. We note that the performance of proposed scheduler is significantly improved ($\Delta V=0.60$) compared to Fig.~\ref{vary_ed1_noPenalty} ($\Delta V=0.33$) wherein $\delta_1$ was arbitrarily set to $b_1 + (4/a_1)$. This is because selecting optimal\footnote{Here we refer to optimality over reduced search space for $\delta_1$.} $\delta_1$ prevents latency of ED class $2$ from growing very large when $\lambda_{\text{ed}}^1$ becomes very large relative to $\lambda_{\text{ed}}^2$.

\subsection{Impact of penalizing PU failures}
\begin{figure}[!t]%
\centering
\includegraphics[scale = 0.5]{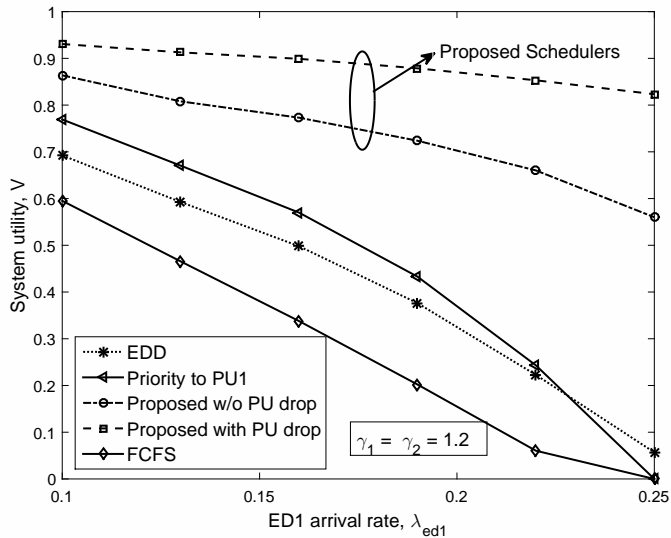}%
\caption{Utility performance for small but equal PU failure penalty.}%
\label{vary_ed1_moderate}%
\end{figure}

\begin{figure}[!t]%
\centering
\includegraphics[scale = 0.5]{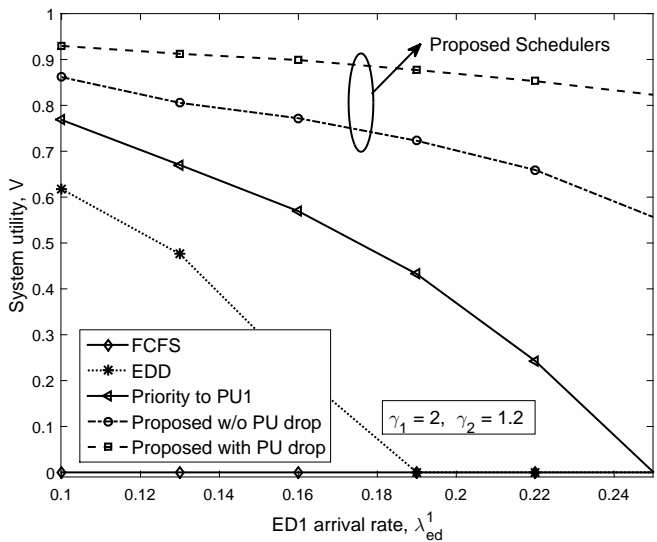}%
\caption{Utility performance for unequal PU failure penalty.}%
\label{vary_ed1_extremePenalty}%
\end{figure}

 We now consider the impact of penalizing PU failures on the performance improvement of different schedulers. Fig.~\ref{vary_ed1_moderate} shows the plot of system utility when the penalty factor is small but equal for both PU classes $(\gamma_1=\gamma_2=1.2)$. We observe that while the utility for other schedulers take a performance hit and approach $0$ at $\lambda_{\text{ed}_1}=0.25$, the proposed scheduler is fairly robust to the penalization. This is because it possesses multiple degrees of freedom in the form of variable latency thresholds which gets re-calibrated when we begin penalizing PU failures. However, other schedulers lack this flexibility and hence their performance suffer. Further increase in penalty of PU class $1$ worsens the performance of other schedulers as shown in Fig.~\ref{vary_ed1_extremePenalty}. But once again, the proposed scheduler is robust and its performance does not decrease noticeably.

\section{Conclusions}
\label{concl}
In this paper, we presented a online delay-efficient multiclass scheduler for delay-heterogeneous M2M uplink traffic. The data from each sensor is classified into PU and ED traffic. Furthermore, due to the delay-heterogeneity across the sensors, the aggregated traffic at M2M application server is classified into multiple PU and multiple ED classes. We use step and sigmoidal functions to represent the different utility of service to PU and ED classes. The average utility of each class is finally aggregated into a proportionally-fair system utility metric. The proposed delay-efficient scheduler uses heuristics that aim to maximize system utility metric. Specifically, it exploits the \emph{firm} utility for PU packets by prioritizing service to ED data as long as we meet the deadline for the PU data. However with increase in network size, an increasingly number of PU packets fail to meet their deadline. We remove the failed PU packets to reduce congestion and improve overall system utility. We also introduce a novel penalty function as part of PU utility to reduce burst PU failures for critical applications. Using extensive simulations, we did a comprehensive delay-performance analysis of the proposed scheduler with respect to other schedulers. We note that the proposed scheduler outperforms other schedulers with the performance gap increasing with heterogeneity in latency requirements and higher penalty for PU failures.

\bibliographystyle{ieeetr}	

\bibliography{wiOptMultiClass}	

\end{document}